\documentclass[aps,pra,preprintnumbers,preprint,amsmath,amssymb]{revtex4}
\usepackage{graphicx,epsfig,epsf}% Include figure files
\usepackage{dcolumn}% Align table columns on decimal point
\usepackage{bm}% bold math

\begin{document}
\newcommand{\ri}{{\rm i}}
\newcommand{\re}{{\rm e}}
\newcommand{\bx}{{\bf x}}
\newcommand{\by}{{\bf y}}
\newcommand{\bd}{{\bf d}}
\newcommand{\br}{{\bf r}}
\newcommand{\bk}{{\bf k}}
\newcommand{\bv}{{\bf v}}
\newcommand{\bw}{{\bf w}}
\newcommand{\bE}{{\bf E}}
\newcommand{\bR}{{\bf R}}
\newcommand{\bM}{{\bf M}}
\newcommand{\bI}{{\bf I}}
\newcommand{\bn}{{\bf n}}
\newcommand{\bs}{{\bf s}}
\newcommand{\tbs}{\tilde{\bf s}}
\newcommand{\rSi}{{\rm Si}}
\newcommand{\beps}{\mbox{\boldmath{$\epsilon$}}}
\newcommand{\rg}{{\rm g}}
\newcommand{\tr}{{\rm tr}}
\newcommand{\xmax}{x_{\rm max}}
\newcommand{\ra}{{\rm a}}
\newcommand{\rx}{{\rm x}}
\newcommand{\rs}{{\rm s}}
\newcommand{\rP}{{\rm P}}
\newcommand{\up}{\uparrow}
\newcommand{\down}{\downarrow}
\newcommand{\hc}{H_{\rm cond}}
\newcommand{\kb}{k_{\rm B}}
\newcommand{\cI}{{\cal I}}
\newcommand{\tit}{\tilde{t}}
\newcommand{\cE}{{\cal E}}
\newcommand{\cC}{{\cal C}}
\newcommand{\Ubs}{U_{\rm BS}}
\sloppy

\title{Indirect decoherence in optical lattices and cold gases} 
\author{Daniel Braun}
\affiliation{Laboratoire de Physique Th\'eorique, IRSAMC, UMR 5152 du CNRS,
  Universit\'e Paul Sabatier, 118, route de 
  Narbonne, 31062 Toulouse, FRANCE} 

%\centerline{\today}
\begin{abstract}
 The interaction of two--level atoms with a common  heat bath leads to an
  effective interaction between the atoms, such that 
  with time the internal degrees of the atoms become correlated or
  even entangled. If part of the atoms remain unobserved this creates
  additional indirect decoherence for the selected atoms, on
  top of the direct decoherence due to the interaction with the
  heat bath. I show that indirect decoherence can drastically increase and
even  dominate the decoherence for sufficiently large times. 
I investigate indirect decoherence through thermal black body radiation
  quantitatively for 
  atoms trapped at regular positions in an optical lattice as well as for
  atoms at random positions in a cold gas, and show how indirect
  coherence can be controlled or even suppressed through experimentally
  accessible parameters.
\end{abstract}
%\pacs{QICS: 02.40.+d}
\maketitle
%\begin{twocolumn}
   
With the rise of quantum information processing it has become
necessary to understand decoherence in true many--particle
systems. It has been known for a
long time that the decoherence rate for a single degree of freedom
scales like a power of a certain distance between the components of
a superposition. The ``distance'' and its natural scale depend on the
coupling to the heat bath. For example, if the single degree of
freedom couples through
a spatial coordinate $x$ to the heat bath, the latter selects
eigenstates of $x$ as ``pointer-basis'' \cite{Zurek81}, and the
relevant ``distance'' is measured in configuration space, with a
microscopic length scale such as the thermal de Broglie length
as natural unit. Decoherence therefore becomes extremely fast for
mesoscopic or even macroscopic distances, and this is considered one of the
main reasons why the everyday world around us behaves
classically. Other couplings lead to different power laws and  
different natural microscopic units \cite{Strunz02}. Decoherence
processes for 
single degree of freedom systems are nowadays routinely resolved
experimentally for microscopic distances between the superposed
components, 
and have been so far in good agreement with the theoretical
predictions
\cite{Brune96,Vion02,Yu02,Hornberger03,Juulsgaard04,Langer05}. 
However, decoherence measurements on true many particle systems are
only now becoming available \cite{KrojanskiS04}, and there is a
need for detailed theoretical predictions, in order to verify the
validity of quantum mechanics in an entirely new regime, namely one
where the joined states of 
many particles are coherently superposed \cite{Duer02,Mintert05,Leggett02}. 

Theoretical progress was achieved
recently with the derivation of a ``decoherence metric''
\cite{Braun06b}, which measures the distance between 
the components of a quantum superposition of arbitrarily many
qubits with degenerate energy levels, and determines directly the
time dependent decoherence. It turned out that for sufficiently far
separated qubits with degenerate energy levels the time dependent
decoherence boils down to just single qubit 
decoherence multiplied with the standard Hamming distance between the
superposed quantum code words. For smaller qubit separations (the relevant
length scale 
is the inverse of the wave length of the UV cut-off of the bath
modes), interference effects start to play a role and one sees strong
deviations from the simple scaling with the standard Hamming
distance. Nevertheless, the notion of a distance 
(more precisely: a pseudo--metric, in the strict mathematical
sense), can be maintained through the introduction of a metric tensor
determined by the heat bath, whose
off-diagonal elements reflect the 
interference processes. As a consequence, $2^{N(N-1)/2}$ independent
decoherences are governed by only $\sim N^2$ matrix elements of the metric
tensor. The
entanglement of a state alone does not determine how fast it decoheres: For
example, a GHZ state $(|000\rangle+|111\rangle)/\sqrt{2}$ has maximum
Hamming distance between its two components, but if the qubits are
sufficiently close the decoherence metric will distinguish this state for
example from the state $(|001\rangle+|110\rangle)/\sqrt{2}$, or all other
states which differ from GHZ by flipping qubits, whereas all theses states
have the same entanglement.

The metric tensor contains the contribution from the direct
decoherence process discussed so far, but also an ``indirect
decoherence'': the heat bath generates effective interactions
between the qubits which can lead to classical correlations or
even entanglement between them (``reservoir induced entanglement'',
\cite{Braun02}). ``Indirect decoherence'' is the additional decoherence
that is induced if some of the atoms which got correlated or entangled with the
selected atoms remain unobserved. In this paper I investigate 
indirect decoherence in more detail, and show that even for a rather small
number 
of unobserved atoms (of the order of 10) and for weak effective coupling
between the atoms, indirect decoherence can
drastically enhance the overall decoherence. The effect
should be important if
one wants to build a quantum memory from
trapped atoms in an optical lattice, but when additional atoms get trapped
in the optical lattice 
and are not read out. Indeed, until recently
\cite{Bloch05} it was difficult 
to even control the number of atoms per lattice site. The situation might be
even worse for quantum information stored in macroscopic gas samples
\cite{Juulsgaard04}, where 
the total number of atoms in which the information is stored can only be
estimated and one has no control over which individual atoms store the
quantum information. 

\section{The Model}
Let us consider $N$ two level atoms at
arbitrary but fixed positions $\bR_i$ ($i=0,\ldots N-1$) interacting with
thermal black body radiation, which forms a common heat bath. All atoms are
assumed identical with degenerate  
energy eigenstates $|-1\rangle$ and $|1\rangle$, with $\sigma_z|\pm 1\rangle=\pm |\pm
1\rangle$. 
In dipole coupling approximation, the total hamiltonian reads \cite{Scully97}
\begin{eqnarray}
H&=&\sum_k \hbar\omega_k a_k^\dagger
a_k+\hbar
\sum_k\sum_{i=0}^{N-1}g_k^{(i)}\sigma_{xi}\left(a_k\re^{\ri
  \bk\cdot\bR_i}+a_k^\dagger\re^{-\ri\bk\cdot\bR_i}\right)\,, \label{H}
\end{eqnarray}
where $\sigma_{xi}$ and $\sigma_{zi}$ are Pauli matrices for atom $i$.
 The index $k$
stands for wave  
vector $\bk$ and polarization 
direction $\lambda$ ($k_j=2\pi n_j/L $ with 
integer $n_j$, $j=x,y,z$ for periodic boundary conditions);
 $a_k^\dagger$ ($a_k$) are the creation 
(annihilation) operators for mode $k$  with frequency $\omega_k=c|\bk|$,
 polarization vector $\epsilon_k$, and electric field amplitude
 $\cE=\sqrt{\hbar\omega_k/(2\varepsilon_0 V)}$, where $\varepsilon_0$, $c$,
 and $V$ are the
 dielectric constant
of the vacuum, speed of light, and the quantization volume, respectively.
 The coupling constant of 
 atom $i$ to mode $k$ is denoted by
 $g_k^{(i)}=-\frac{ed\cE}{\hbar}\hat{u}^{(i)}\cdot\epsilon_k$, where 
 $\hat{u}^{(i)}$ stands for a unit vector 
in the direction of the dipole moment of atom $i$, $\langle -1|\bd|1\rangle=ed
\hat{u}^{(i)}$ with electron charge $e$ and dipole length $d$.
The restriction to atoms with degenerate 
energy levels, $\Omega_0=0$, leads to a vanishing system hamiltonian, $H_{
\rm sys}= \frac{1}{2}\hbar\Omega_0\sum_{i=0}^{N-1}\sigma_{zi}=0$. The model is
 a special case of the more general class of models known as pure
 dephasing models, where the system hamiltonian commutes with the
 interaction hamiltonian, i.e.~the second term in (\ref{H}). These models
 can be solved exactly  
for an arbitrary number of atoms at arbitrary positions. 

\section{Decoherence metric}
We are interested in the decoherence process of the $n$ selected atoms (indices
$0,\ldots,n-1$) out of the $N$ atoms. We therefore have to first trace out
the electro--magnetic (e.m.) field 
modes, leaving a density matrix $\rho$, and secondly the unobserved atoms
$n\ldots N-1$. The resulting reduced 
density matrix $\tilde{\rho}(t)$ of the remaining atoms will
be expressed in the eigenbasis of the $\sigma_{xi}$,
$\sigma_{xi}|\pm\rangle_x=\pm|\pm\rangle_x$, the natural basis (also
called pointer basis) for
studying the decoherence process \cite{Zurek81}. It has
matrix elements 
$\tilde{\rho}_{\tbs\tbs'}(t)=\tr_{n\ldots N-1}\rho_{\bs\bs'}(t)$, where
$\tbs$ and $\tbs'$ are subsets of length $n$ of the labels 
$\bs=(s_0,s_1,\ldots
s_{N-1})$ and $\bs'=(s_0',s_1',\ldots s_{N-1}')$ of 
the quantum states $|{\bf s}\rangle$ and $|{\bf
  s}'\rangle$ of all atoms, taken as column vectors, and $s_i,s_i'=\pm 1$,
$i=0,1,\ldots,N-1$, refer to atom $i$. We assume that all unobserved atoms
are initially in the energy eigenstate
$|1\rangle=(|1\rangle_x+|-1\rangle_x)/\sqrt{2}$, and that there are no
initial correlations between the unobserved atoms and the selected atoms.

The dynamical quantities of interest are the ``decoherences''
$d_{\tbs\tbs'}(t)$, which we define as normalized complements of
``coherences''  
(i.e.~off-diagonal elements of the reduced density matrix of the $n$ selected
atoms alone),
\begin{eqnarray}
d_{\tbs\tbs'}(t)&\equiv&1-\frac{|\tilde{\rho}_{\tbs\tbs'}(t)|}{|\tilde{\rho}_{\tbs\tbs'}(0)|} 
\mbox{ for }\tilde{\rho}_{\tbs\tbs'}(0)\ne 0\,.
\end{eqnarray}
In \cite{Braun06b} it was shown that the behavior of the decoherences
is given by $d_{\tbs\tbs'}(t)\simeq||\tbs-\tbs'||_{M(t)}^2$ with the
``decoherence metric''
\begin{equation} \label{dM}
||\tbs-\tbs'||_{M(t)}
  \equiv\frac{1}{2}\sqrt{(\tbs-\tbs')^T\bM(t)(\tbs-\tbs')}\,,   
\end{equation}
where $^T$ denotes the transpose, and $\bM(t)$ is a real, symmetric, and
non--negative time dependent ``decoherence metric tensor'' (DMT) with matrix
elements ($i,j=0,\ldots,n-1$, $\beta=1/k_BT$ is the inverse temperature), 
\begin{eqnarray}
M_{ij}(t)&=&4f_{ij}(t,\bR_i-\bR_j)+2\Phi_{ij}(t,\bR_i,\bR_j)\,,\label{Mij}\\
f_{ij}(t,\bR)&=&\sum_k\frac{g_k^{(i)}g_k^{(j)}}{\omega_k^2}\cos(\bk\cdot\bR)
(1-\cos\omega_kt)\coth\frac{\beta\hbar\omega_k}{2}\,\label{fij}\\
\varphi_{ij}(t,\bR)&=&2\sum_k\frac{g_k^{(i)}g_k^{(j)}}{\omega_k^2}\cos(\bk\cdot\bR)(\omega_kt- 
\sin\omega_kt)\label{phij}\,\\  
\Phi_{ij}(t,\bR_i,\bR_j)&=&\sum_{k=n}^{N-1}
\varphi_{ik}(t,\bR_i-\bR_k)\varphi_{jk}(t,\bR_j-\bR_k)\label{Phij}\,.
\end{eqnarray}
The heat--bath itself therefore
induces a natural distance $||\tbs-\tbs'||_{M(t)}$ between the $n$--qubit
states, 
which determines directly the time dependent decoherences. The validity of
eq.(\ref{dM}) is limited to $|M_{ij}(t)|\ll 1\,\,\,\forall i,j$.

The distance $||\tbs-\tbs'||_{M(t)}$ generalizes the well--known Hamming
distance 
$D^H(\tbs,\tbs')$, which is defined as the number of 
bits in which $\tbs$ and $\tbs'$ differ, and which is obtained for
$\bM=\bI$. This limit is reached for 
sufficiently large separation of the qubits, and $d_{\tbs\tbs'}(t)$ then
goes over 
into $D^H(\tbs,\tbs')$ up to a time dependent function
describing single qubit decoherence \cite{Braun06b}. 

It is clear from the definition that $\bM(t)$ is real and
symmetric.  In
appendix \ref{appA}  I show that $\bM(t)$ is also non--negative and obeys the
triangle 
inequality. However, if a decoherence free
subspace (DFS) \cite{Zanardi97,Duan98,Lidar98,Braun98,Altpeter04} exists,
the decoherence metric is strictly speaking a pseudo--metric, as there can be
code words $\tbs$ and $\tbs'$ with $\tbs\ne\tbs'$ such  
that $||\tbs-\tbs'||_{M(t)}=0$.

\section{One selected atom}
Indirect decoherence, i.e.~the part $\Phi_{ij}$ in eq.(\ref{Mij})
is best appreciated 
for just $n=1$ selected atom (taken to have index $i=0$), and $N-1$
non--observed 
atoms. Only one decoherence is then relevant, $d_{1-1}$, and $\bM$
has a single matrix element, $M_{00}(t)=4f_{00}(t)+2\Phi_{00}(t)$. 
We express all lengths in terms
of the dipole length $d$, $r_{ij}=|\bR_i-\bR_j|/d$, and times in units 
of $d/c$. We will 
furthermore assume that all dipoles are oriented in the same direction
$\hat{u}$. The 
angle between $\hat{u}$ and the vector $\br_i-\br_j$ will be called
$\theta_{ij}$, such that $\varphi_{ij}(t,\bR_i-\bR_j)$ becomes a function of
$t$, $r_{ij}$, and $\theta_{ij}$, and depends on $i$ and $j$ only through
these variables,
$\varphi_{ij}(t,\bR_i-\bR_j)\equiv\varphi(t,r_{ij},\theta_{ij})$. 
As shown in \cite{Braun06b}, $f_{00}$ is given for $T=0$ by
\begin{equation} \label{fii}
f_{00}(t)=\frac{2}{3\pi}\alpha\left(\frac{\kappa^2}{2}+\frac{1-\cos(\kappa
t)-\kappa t\sin(\kappa t)}{t^2}\right) \,,
\end{equation}
where $\kappa=k_{\rm max}d$ is a UV cut--off of the heat bath. A necessary
condition for the dipole-coupling approximation is $\kappa\ll 1$.
Corrections to (\ref{fii}) due to 
finite temperature are of order $k_BT/(\hbar\omega_{\rm max})$ with
$\omega_{\rm max}=c k_{\rm max}$, and will be
neglected in the following. Eq.(\ref{fii}) implies that for a finite
UV--cutoff the direct decoherence will remain finite for all times, 
\begin{equation} \label{fiil}
f_{00}\stackrel{t\gg 1}{\longrightarrow}\frac{\alpha\kappa^2}{3\pi}\,.
\end{equation}
 Such a behavior has been termed ``incomplete decoherence''
 \cite{Braun02}. The initial behavior is 
quadratic in $t$, $f_{00}\simeq \gamma^2 t^2$ with
$\gamma=\sqrt{\alpha/(12\pi)}\kappa^2$ for $\gamma t\ll 1$. 

The function $\varphi(t,r,\theta)$ reads 
\begin{eqnarray}
        \varphi(t,r,\theta)&=&\frac{2 \alpha}{\pi r^2}
\Bigg\{
  \frac{1}{4}\left(1 + 3 \cos(2 \theta)\right) 
          \Big( \rSi((r + t) \kappa) -
            \rSi((r - t) \kappa )\Big)\nonumber\\
&&     +  \frac{3 \sin^2\theta -2}{2 r}
\Big(
            (r + t)  
             \rSi((r + t) \kappa) - 
            (r -t) 
             \rSi((r - t) \kappa)-2 t \rSi(\kappa r)
\Big)
\Bigg\}\,.
\end{eqnarray}
Rapidly oscillating terms of the type $\sin(\kappa r)$, $\cos(\kappa r)$,
and $\kappa r\cos(\kappa r)$ have been neglected here, as their average in
the case of a small
uncertainty in the atom positions is
exponentially small: atoms of mass $M$ trapped in the ground states of harmonic
oscillators with trapping frequency $\nu$ have a Gaussian distribution of
their center of mass with a 
width $\delta r\sim \sqrt{\hbar/\nu M}$, leading to a suppression of these
terms by a factor $\exp(-(\kappa\delta r)^2/2)$. A typical experimental
parameter,
$\nu\sim$ 30kHz \cite{Greiner02}, leads to $\delta r d\sim 100$nm. Optical
dipole lengths $d\sim${\AA} and a  UV cut-off $k_{\max}\sim$
  1/10{\AA} give $\kappa\sim 0.1$, and $\kappa \delta r\sim 100$, so that
    these terms can indeed be safely neglected. For a cold gas $\delta r$ is
    expected to be even larger.  

In the limit of $|r\pm t|\kappa\gg 1$, $\varphi_{ij}(t,r,\theta)$ approaches
\begin{equation} \label{phijh}
\varphi(t,r,\theta)=\alpha \frac{t}{r^3}(3\cos^2\theta-1)\Theta(t/r-1),
\end{equation}
where $\Theta(t/r-1)$ is the Heaviside function centered on the light
cone. We therefore recognize $\varphi$ as a phase accumulated due to an
effective dipole interaction between the atoms mediated through the modes of
the electromagnetic field. 
Note that in this limit indirect decoherence becomes basically
independent of the cut--off $\kappa$. 

\subsection{Optical lattices}
In the following we consider specifically the situation for a 2D square
optical lattice with lattice constant $a$ (taken in units of the dipole
length $d$ as well), with the single selected atom at the center of the
lattice. 
The fact that 
$\varphi(t,r,\theta)\propto 
t$ leads to an unbound quadratic growth of $\Phi_{ij}(t,\bR_i,\bR_j)$
with $t$, $\Phi_{ij}\sim N_{nn}(\alpha t/a^3)^2$, where $N_{nn}$ is an
effective number of nearest neighbors, weighed by the inverse cube of their
distance from the selected atom in units of the lattice spacing. As
$\alpha\kappa^2\ll 1$ an immediate 
consequence of 
eqs.(\ref{phijh},\ref{fiil}) is 
that for large enough times, $t>t_1$ with 
\begin{equation} \label{ti}
t_1\sim \frac{\kappa a^3}{\sqrt{3\pi\alpha N_{nn}}}\,,
\end{equation}
indirect decoherence always dominates over
direct coherence  even for a small
number of unobserved atoms close  to the 
selected atom. Due to the strong 
$r$-dependence of (\ref{phijh}), the nearest neighbors and next nearest
neighbors give the by far leading contributions to the indirect decoherence. 
Indirect decoherence dominates immediately (i.e.~as soon as $t\gg a\gg 1$,
where 
the quadratic behavior of $\Phi_{ij}\sim N_{nn}(\alpha t/a^3)^2$ in $t$
is valid) over
direct decoherence, if $N_{nn}(\alpha t/a^3)^2\gg\gamma^2 t^2$, or $a<a_c$
with the critical spacing $a_c\sim
(12 \pi\alpha N_{nn})^{1/6}/\kappa^{2/3}$.
If the cut-off is of the order $\hbar\omega_{\rm max}\sim 1$eV and $d=1$\AA,
$a_c d$ 
is of the order 100nm.
If the cut-off is given by the break--down of the dipole
approximation ($k_{\rm max}\sim 2\pi/d$), $a_c$ reduces to
$a_c\sim 1$, i.e. the atoms will have to become basically closely
packed before 
indirect decoherence immediately dominates over direct decoherence. 

\begin{figure}
\epsfig{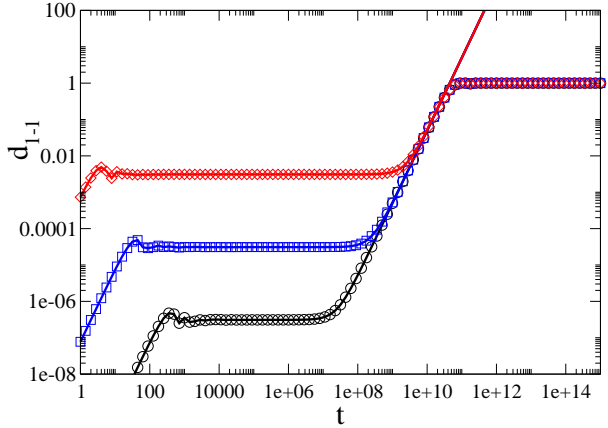}
\caption{The decoherence $d_{1-1}(t)$ for a single selected atom in the
  center of a 2D square optical lattice of 31x31 atoms with lattice constant
  $a=1000$ (in units of the dipole length) as a function of the dimensionless
  time $t$ for 
  $\kappa=0.01$ (black circles), $\kappa=0.1$ (blue squares), and $\kappa=1$
  (red diamonds) along with the decoherence metric predictions (continuous
  lines). The first rise corresponds to the contribution of direct
  decoherence due to the interaction with the e.m.~modes, the second rise
  results from indirect decoherence due to the effective interaction with
  the non-selected atoms mediated by the e.m.~modes. Superpositions are
  decohered completely when $d_{1-1}=1$ is reached. \\[0.2cm]} 
\label{fig.d}       
\end{figure}

One might object that a small
number of two--level atoms with which the selected atoms effectively
interact cannot constitute a 
real heat-bath, and should rather lead to repeating revival phenomena
of the 
coherences instead of to decoherence. However, it turns out that even for
a square optical lattice of 3$\times$3 atoms (i.e.~8 unobserved atoms),
the revivals are hardly visible, and in a square optical lattice of 
31x31 atoms all revivals seem to have disappeared completely. 
This is shown in  
Figure \ref{fig.d} where we see the decoherence as function of time 
for a 2D square optical lattice with lattice constant $a=1000$, dipole 
moments perpendicular to the plane of the 
lattice, and with the selected atom in the center of the lattice. Direct
decoherence sets 
in immediately and increases $\propto t^2$ for small times (see
eq.(\ref{fii})), before saturating at $\alpha\kappa^2/(3\pi)$. Due to the
weakness of the dipole coupling, indirect decoherence becomes appreciable
only at much later times for atoms separated thus far.  But because of 
the continued quadratic growth of the indirect decoherence it finally
destroys all coherence left by the direct decoherence.
The figure also shows that the exact
result for $d_{1-1}$ \cite{Braun06b} is very well approximated in the
entire interesting 
regime $d_{1-1}\le 1$ by the decoherence metric
prediction, eq.(\ref{dM}).  

Indirect decoherence has an interesting dependence on the orientation of the
dipoles. Eq.(\ref{phijh}) shows that
$\Phi_{00}\propto(3\cos^2\theta-1)^2$. Indirect decoherence in a 2D optical
lattice can therefore be
completely suppressed by orienting the atomic dipoles at the magical angle
$\theta=\arccos(1/\sqrt{3})\simeq 
54.7^o$ with respect to the lattice. This angular dependence might serve as
additional experimental signature of the effect.  

\subsection{Cold gases}

For an atomic gas, the positions $\bR_j$ of the atoms are not
known. We resort to an ensemble description, where we average over the
positions of the 
atoms. We assume that all $\bR_j$ with the exception of $\bR_0={\bf 0}$ are
randomly, independently and evenly 
distributed with an average density (atoms per volume) $\rho_V$. We find
\begin{eqnarray} \label{avphi00}
\langle \Phi_{00}(t,{\bf 0},{\bf 0})\rangle&=&\langle
  \sum_{k=1}^{N-1}\varphi_{0k}^2(t,\bR_k)\rangle\nonumber\\
&=&2\pi\rho_V d^3\int_l dr\,r^2\int_0^\pi d\vartheta \sin\vartheta (\alpha
t/r^3)^2(3\cos^2\vartheta -1)^2\Theta(t/r-1)\,.
\end{eqnarray}
The lower cut-off is now given by the smallest distance up to which two
 atoms might approach each other, which for a sufficiently dilute gas at low
 temperature is of the order of the scattering
 length $l$ (taken in units of $d$ as well), if $l>0$. This gives 
\begin{equation} \label{phi00g}
\langle 
 \Phi_{00}(t,0,0)\rangle\simeq \gamma_G^2 t^2 \mbox{ with }\gamma_G=\alpha 
 \sqrt{\frac{16\pi d^3\rho_V}{15l^3}}\,.
\end{equation}
 Indirect decoherence immediately dominates over direct decoherence for
 densities 
 $\rho_V\gtrsim \kappa^4 l^3/d^3$. A
 UV cut-off $\hbar\omega_{\rm max}=1$eV, $d=1${\AA}, $l=10${\AA} gives a
 critical  
 density of about $10^{20}$ atoms/m$^3$. Recent experiments on dense
 Bose-Einstein condensates deal already with similar densities
 \cite{Schuster01}. The coherence of internal (spin) 
 degrees of 
 freedom of condensed bosons has recently been demonstrated \cite{Chang05},
 so that indirect decoherence in a cold gas might become observable in the
 near future. 
For  smaller densities 
 indirect decoherence takes over for $t>t_2$ with  
\begin{equation} \label{t2}
t_2\sim\kappa\sqrt{\frac{l^3}{\rho_Vd^3}}\,.
\end{equation}
The dependence of the indirect decoherence on $l$ gives the interesting
perspective to control indirect decoherence in a cold gas through a Feshbach
resonance, which allows to vary $l$ over many orders of magnitude
\cite{Feshbach58}. 

\section{Conclusions}
I have shown that indirect decoherence due to reservoir induced entanglement
between  degenerate two--level atoms
can substantially increase decoherence in an optical lattice or a
cold atomic gas, compared to the direct decoherence due to the 
coupling of each atom to the e.m.~field. For large enough times indirect
decoherence in fact always 
dominates, even for only a few unobserved atoms. For
sufficiently densely packed atoms
the dominance of indirect decoherence begins as
soon as a light signal has traveled a dipole length. The dependence of the
indirect decoherence on the orientation of and the distance between the dipoles
offers the interesting perspective to control indirect decoherence with
easily accessible parameters. In a 2D optical lattice, indirect decoherence
can be switched off completely by orienting all dipoles under a magical
angle $\theta=\arccos(1/\sqrt{3})$ with respect to the lattice, and in a
dilute, cold atomic gas, one can suppress indirect decoherence to large
extent by increasing the scattering length $l$ through a Feshbach resonance.

\begin{acknowledgments}
I wish to thank Olivier Giraud for useful discussions. This work was
supported in part by the Agence 
National de la Recherche 
(ANR), project INFOSYSQQ, and EC IST-FET projects EDIQIP and EuroSQIP.
\end{acknowledgments}

\section{Appendix A: Proof of non--negativity and of the triangle inequality
  for the decoherence 
  metric}\label{appA} 
\subsection{Non--negativity}
We show separately $\sum_ix_if_{ij}x_j\ge0$ and $\sum_ix_i\Phi_{ij}x_j\ge0$
$\forall 
x_i\in\mathbb{R}$. From eqs.(\ref{fij},\ref{phij}) we have\\
\begin{eqnarray}
\sum_{i,j}x_i f_{ij} x_j&=&\sum_{i,j}\sum_k x_i x_j
g_k^{(i)}g_k^{(j)}\re^{i\bk\cdot(\bR_i-\bR_j)}\frac{1-\cos\omega_k
t}{\omega_k^2}\coth\frac{\beta\hbar \omega_k}{2}\nonumber\\
&=&\sum_k\left|\sum_ix_ig_k^{(i)}\re^{\ri\bk\cdot
  R_i}\right|^2\frac{1-\cos\omega_k
t}{\omega_k^2}\coth\frac{\beta\hbar\omega_k}{2}\ge
0\nonumber\\
\sum_{i,j}x_i \Phi_{ij}
x_j&=&\sum_{i,j}\sum_lx_i\varphi_{il}\varphi_{jl}x_j=\sum_l\left(\sum_ix_i\varphi_{il}\right)^2\ge 0\,.\nonumber
\end{eqnarray}
Thus, also $\sum_{i,j}x_iM_{ij}x_j\ge0$, and $\bM$ is therefore non--negative.
\subsection{Triangle inequality}
We define the linear map $\phi: \,\,\,\mathbb{R}^n\to\mathbb{R}^n$, $\bv\to \bM
\bv$, 
where $\bM$ is a real symmetric, non--negative $n\times n$ matrix,
i.e.~$\bv^T\bM\bv\ge 0$ for all $\bv\in \mathbb{R}^n$. We also define the
bi-linear form 
$(\cdot,\cdot):\mathbb{R}^n\times\mathbb{R}^n\to\mathbb{R}$,
$(\bv,\bw)=\bv^T\bM\bw$, which is not a  scalar product, as $(\bv,\bv)$ can
be zero for $\bv\ne 0$. One can nevertheless prove the Cauchy---Schwartz
(C.S.)  
inequality $(\bv,\bw)^2\ge(\bv,\bv)(\bw,\bw)$:

Let $V_0$ be the kernel of $\phi$. Thus $\bM\bv=0=\bv^T\bM\,\,\,\,\forall \bv\in
V_0$. Suppose 
first that $\bv\in V_0$ or $\bw\in V_0$. Then $(\bv,\bw)=0$, but also
$(\bv,\bv)(\bw,\bw)=0$, as at least one factor is zero. Thus the
C.S.~inequality is trivially fulfilled. Now suppose that $\bv\notin V_0$
and $\bw\notin V_0$. Define $\tilde{\bv}=(\bw,\bw)\bv-(\bw,\bv)\bw$. We
have 
\begin{equation} \label{vv}
0\le(\tilde{\bv},\tilde{\bv})=(\bw,\bw)[(\bw,\bw)(\bv,\bv)-(\bw,\bv)^2]\,.
\end{equation}
 It is
easily seen that $(\bw,\bw)\ne 0$ if $\bw\notin V_0$: Decompose
$\bM=\bM_0+\bM_+$ 
with $\bM_0=P_0\bM P_0$, $\bM_+=(1-P_0)\bM(1-P_0)$, where $P_0$ is the
projector onto 
$V_0$. $\bM_+$ is the positive part of the map,
i.e.~$\bw^T\bM_+\bw>0\,\,\,\forall 
\bw\ne 0$. Thus, from $(\bw,\bw)=0$ follows $\bw=0$ or $\bM_+=0$. In both cases
$\bw\in V_0$. Thus, for $\bw\notin V_0$ we have $(\bw,\bw)\ne 0$, and as
$\bM\ge 0$, this means $(\bw,\bw)> 0$. Therefore eq.(\ref{vv}) immediately
gives the C.S.~inequality. 

The proof of the triangle inequality
$||\tbs-\tbs''||_M\le||\tbs-\tbs'||_M+||\tbs'-\tbs''||_M$ then proceeds in the
usual fashion. One 
defines the norm $||\bx||_M=\sqrt{(\bx,\bx)}$, and the C.S.~inequality gives
$||\bx+\by||_M^2=(\bx,\bx)+(\by,\by)+2(\bx,\by)\le
(\bx,\bx)+(\by,\by)+2|(\bx,\by)|\le 
||\bx||_M^2+||\by||_M^2+2||\bx||_M\,||\by||_M=(||\bx||_M+||\by||_M)^2$. The
triangle 
inequality for the distance (\ref{dM}) follows from here
by setting $\bx=\tbs-\tbs'$, $\by=\tbs'-\tbs''$.
\bibliography{../mybibs_bt}
%\begin{thebibliography}{im4}
%\end{thebibliography}
%\end{twocolumn}
\end{document}